\documentclass[manyauthors]{fundam-stef}
\input{stefmac.sty}
\usepackage{url,listings,graphics,graphicx,color,latexsym,amssymb,amsmath,amsfonts,verbatim,rotating,epsfig}
\usepackage[all]{xy}

\begin{document}
\setcounter{page}{1}
\issue{}

\title{Adaptive virtual organisms: A compositional model for complex hardware-software binding\thanks{Parts of the results in this paper have been presented to the FROM-2018 conference, Iasi, Romania, June 13-15, 2018.}\thanks{The research presented here has been partially supported by the Data Science Research Center, University of Bucharest and by the ATLAS project (Id, 2018-2020).}}
\author{Ciprian Ionut Paduraru \\ Department of Computer Science\\ University of Bucharest, Romania\\ciprian.paduraru{@}fmi.unibuc.com
\and Gheorghe Stefanescu\\ Department of Computer Science\\ University of Bucharest, Romania\\ gheorghe.stefanescu{@}fmi.unibuc.ro} 
\maketitle
\address{{\em C.I.~Paduraru \&\ G.~Stefanescu, Department of Computer Science, University of Bucharest, Str Academiei 14, Bucharest 010014, Romania}}
\runninghead{C.I.~Paduraru, G.~Stefanescu}{Adaptive virtual organisms}

\begin{abstract}
The relation between a structure and the function running on that structure is of central interest in many fields, including computer science, biology (organ vs. function), psychology (body vs. mind), architecture (designs vs. functionality), etc. Our paper addresses this question with reference to computer science recent hardware and software advances, particularly in areas as robotics, AI-hardware, self-adaptive systems, IoT, CPS, etc. 

At the modeling, conceptual level, our main contribution is the introduction of the concept of ``virtual organism'' (VO), to populate the intermediary level between rigid, slightly reconfigurable, hardware agents and abstract, intelligent, adaptive software agents. A virtual organism has a structure, resembling the hardware capabilities, and it runs low-level functions, implementing the software requirements. Roughly speaking, it is an adaptive, reconfigurable, distributed, interactive, open system, consisting of a network of heterogeneous computing nodes, with a constrained structural shape, and running a bunch of overlapping functions. The model is compositional in space (allowing the virtual organisms to aggregate into larger organisms) and in time (allowing  the virtual  organisms to get composed functionalities).

Technically, the virtual organisms studied here are in 2D (two dimensions) and their structures are described by regular 2D pattens; adding the time dimension, we conclude our VO model is in 3D (with 2D for space and 1D for time). By reconfiguration, an organism may change its structure to another structure belonging to the same 2D pattern. We illustrate the VO concept with three increasingly more complex VOs: (1) a tree collector organism (TC-organism); (2) a feeding cell organism (FC-organism); and (3) an organism consisting of a collection of connected feeding cell organisms. 
To test the benefits of reconfiguration, we implemented a simulator for TC-organisms and the quantitative results confirm the intuition that reconfigurable structures are better suited than fixed structures in dynamically changing environments.

Finally, we briefly show how Agapia may be used for getting quick implementations for VO's simulation. Agapia is a structured parallel, interactive programming language where dataflow and control flow structures can be  freely  mixed.  Currently,  its  compiler  produces  HPC  runs,  within  either  MPI  or  OpenMP environments.
\end{abstract}



\section{Introduction}

The relation between a structure and the function running on that structure is of central interest in many fields, including computer science, biology (organ vs. function), psychology (body vs. mind), architecture (designs vs. functionality), etc. Our paper addresses this question with reference to computer science recent hardware and software advances. We recall the structure-function relation in the simple case of single-processor and sequential-programs, then we look at possibilities to lift this relation to today complex hardware and software systems.

\subsection{Background}

In computer science, the structure-function distinction occurs  as the hardware-software dichotomy. Traditionally, hardware and software worlds are tied together by an unwritten pact: the Instruction Set Architecture (ISA). The hardware companies develop  small, powerful, less energy consuming computers, implementing the ISA instructions in hardware. On the other side, the software companies develop efficient, reliable, user-friendly programs, together with compilers and related needed tools to translate them into ISA instructions. The ISA pact is essential for running the software on the hardware. These fields are separated, but they  are cross-influencing each other: techniques form software are adopted by hardware producers (e.g., pipeline, multi-threading), and conversely, hardware advances may lead to development of new software fields (e.g., applications in CUDA).

Recent hardware and software advances, particularly in areas as IoT (the Internet of Things), robotics, self-adaptive systems, CSP (Cyber-Physical Systems), AI-hardware, etc. have lead to  heterogeneous environments where the structure-function distinction is not easily recognized. The today new hardware is a conglomerate of connected devices, often requiring specialized software to integrate them. Similarly, today complex software applications require integration of various functions, supported by different platforms, sometimes requiring specific hardware support (e.g., networking applications). The achieved old goal, supported by the ISA architecture, to have any program running on any machine, is a far reaching goal again. 

Hence, a the following question naturally arises: 
\begin{quote}
Can we identify a new ``ISA'' for this new, distributed computing setting? What a ``program'', or a ``computer'' running it, actually is in this new context?
\end{quote}

We see an emerging trend coming from both directions and consolidating something in the middle, named in this way: \textit{middleware}. On the one hand, there is an increasing support for software composition, via services and micro-services, going down to a finer and better controlled granularity. On the other hand, advances in computing and networking lead to more flexible and better integrated networks of devices. With proper development, these trends may lead to a better portability of software, running across multiple hardware combinations - this is particularly important for the highly heterogeneous IoT applications.  

\textit{Autonomic computing} was promoted by IBM \cite{autonomic2003}, suggesting that complex computing systems should have autonomic properties, to independently take care of the regular maintenance and optimization tasks. IBM has also identified four basic properties of self-managing systems\foo{The Virtual Organism model, presented below, was designed to naturally support these features.}: self-configuration, self-optimization, self-healing, and self-protecting. Current architecture design promotes ``separation of concerns'', considering two different layers: one for the system functionality, the other for its adaptation.

Coming back to the single-processor and sequential-programs case,
we see that the classical ISA pact is based on a nice feature of \textit{regular languages} \cite{kleene1951representation, conway2012regular}. Namely, regular languages, used as a control mechanism here, occur in two equivalent forms: on the one hand, they are recognized by \textit{finite automata} and these automata are well suited for hardware design; on the other hand, they are specified by \textit{regular expressions} and these expressions are essential for modeling control in structured programs and for other further software developments.
Along this line of reasoning, one suggestion is to try to extend regular languages and the finita-automata vs. regular-expressions relationship in 2D (two dimensions) and to see if this model may support a kind of new ISA pact for complex distributed systems.

Studies on extending (regular) languages in 2D started form 1960s. The field of 2D formal languages \cite{giammarresi1997two, lindgren1998complexity, wolfram2002new} is now a mature one, with many applications in various areas, ranging form computer science to biology, from physics to sociology. Our interest in these extensions has roots in models and programming languages for open, interactive, distributed systems, based on space-time duality, with 1D used for space and 1D used for time; a brief survey is presented in \cite{ecsa}. In more recent papers, including \cite{bps13,DBLP:conf/RelMiCS/Stefanescu15,ecsa}, we extended the 2D setting to arbitrary shapes words, particularly introducing new 2D regular expressions. A precursor of the present paper is \cite{DBLP:conf/saso/PaduraruMS17}, which introduces a formal multi-level approach for developing systems with controlled structure and functionality, using two dimensions for space and one for time. Technically, the key new ingredient added in the present paper is adaptation and the resulting neat concept of virtual organisms. 

From this sketched perspective, one goal may be: 
\begin{quote}
To design a compositional computing model, supporting a kind of \emph{distributed assembly language} for modern hardware and software systems, filling the gap between abstract, complex software systems and rigid, reconfigurable, distributed, hardware architectures.
\end{quote}
The computing model and the associated programming language should be, among other: (i) compositional; (ii) expressive and easy to use; (ii) including structural and functional constraints; (iii) allowing for quantitative evaluation of structural reconfiguration cost; (v) allowing for quantitative evaluation of functional reconfiguration cost (adding, removing, or overlapping functions); (vi) including features achievable to hardware designers and expressive enough for developing software applications.

\subsection{Results}

At the modeling, conceptual level, our main contribution is the introduction of the concept of {\em virtual organism}, to populate the intermediary level between rigid, but slightly reconfigurable, hardware agents and abstract, intelligent, adaptive software agents. A virtual organism has a  structure, resembling the hardware capabilities, and it runs low-level functions, implementing the software requirements. Roughly speaking, it is an adaptive, reconfigurable, distributed, interactive, open system consisting of a network of heterogeneous computing nodes with a constrained structural shape and running a bunch of overlapping functions.
The model is compositional in space (allowing the virtual organisms to aggregate into larger organisms) and in time (allowing  the virtual  organisms to get composed functionalities).

Technically, the virtual organisms in this paper are in 3D, with 2D for space and 1D for time, and their structure is constrained by regular 2D pattens. By reconfiguration, an organism may change its structure to another structure belonging to the same 2D pattern. Two classes of reconfigurations are particularly important: conservative reconfigurations (preserving the nodes, but changing the structure) and elastic reconfigurations (allowing for adding or removing nodes).

To illustrate the concept we briefly present three simple, increasingly more complex, virtual organisms: (1) a tree collector organism (TC-organism); (2) a feeding cell organism, consisting of a membrane, with attached collecting and releasing trees (FC-organism); and (3) a collection of connected, feeding cells organisms (CFC-organism).

To test the benefits of reconfiguration, we implemented simulators for TC- and FC- organisms (only the results for TC-organisms are reported here). A TC-organism has a tree structure and collect items from multiple sources with a flow from leafs to root, under the following constraints:  (i) there is a cap on the allowed flow per node; and (ii) leafs' collecting capabilities depend on their distances to sources. 
We tested the following scenarios: (1) for fixed sources, we checked the average number of (conservative) reconfiguration steps needed by a random tree to reach an optimal shape; (2) for variable sources, we compared an initially random, but conservatively reconfigurable, tree collector against a fixed tree collector,  optimal for the initial configuration; (3) finally, we compared conservative and elastic reconfigurations, considering the costs of node rental and of collected items.

The quantitative results have confirmed the intuition that: (1) {\em reconfigurable structures are better suited than fixed structures in dynamically changing environments}; and (2) {\em elastic reconfiguration is better suited than conservative reconfiguration, when the cost of node rental and of unit flow collected are taken into account}.

\paragraph*{Contents of the paper} 
After a brief presentation of regular 2D patterns (Section 2), we introduce virtual organisms (Section 3); then, we describe an implementation for simulating and visualizing virtual organisms' behavior (Section 4); finally, there is a short section on conclusions, related and future work.

\section{Regular 2D patterns}\label{sec2}

As we already emphasized above, we will introduce a computing model (of virtual organisms) at the interface between concrete, reconfigurable hardware and  abstract, adaptive, intelligent software. To this end, we separate the underlined physical structure from the functionalities mapped on that structure. 

Technically, we use regular 2D patterns\foo{Our model is in 2D, but can be extended in higher dimensional spaces. For the sake of generality, we started with regular 2D patterns; however, considering additional variables (enriching border labels) and restrictions, classes of more constrained 2D patterns may be defined.} for organism structure. Their role is to constrain the reconfiguration process: a structure may be replaced by another structure only if the latter belongs to the same 2D pattern. For mapping functionalities, certain low-level functions on the computing nodes (local computations and communications) are used to implement the higher level functionalities, required by the software.   

\subsection{Regular 2D patterns}

In this subsection we define regular 2D patterns (see \cite{DBLP:conf/RelMiCS/Stefanescu15} for more information and additional references) and describe three increasingly more complex regular 2D patterns: \emph{trees}, \emph{rings with attached trees}, and \emph{connected rings with attached trees}. 

A \emph{2D letter} is a labeled unit cell in 2D. A \emph{2D word} is a connected area of 2D letters, where the letters are connected along their west, north, east, or south borders. 

A \emph{2D pattern} (or \emph{2D languages}) is a set of 2D words.  We present three mechanisms for defining 2D patterns\foo{The first two representations are equivalent, provided one adds renaming to (i). On the other hand, currently it is an open question to find a class of regular 2D expressions corresponding to 2D automata, hence to get a Kleene theorem in this 2D setting; see \cite{DBLP:conf/RelMiCS/Stefanescu15}.}: (i) by tiling, using 2D automata; (ii) by product of (two) regular languages; or, (iii) by regular 2D expressions.  

\paragraph{2D automata}
A \emph{2D automaton} consists of: (1) a finite set of tiles (with body and border labels); and (2) a specified set of accepting labels for the external borders. A \emph{word $w$ is accepted} by such an automaton $T$ if there is a tiling, with tiles from $T$, having the selected labels on its external borders, and whose projection on the  body labels is $w$. The language accepted by a 2D automaton $T$ is denoted by $L(T)$. These languages, accepted by 2D automata, are called \textit{regular}.

\begin{figure}
\centerline{
\raisebox{.3cm}{\includegraphics[scale=0.2]{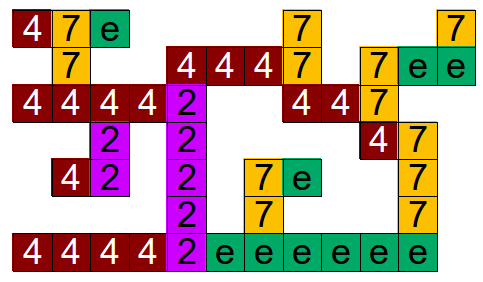}}\hspace{.5cm} \includegraphics[scale=0.2]{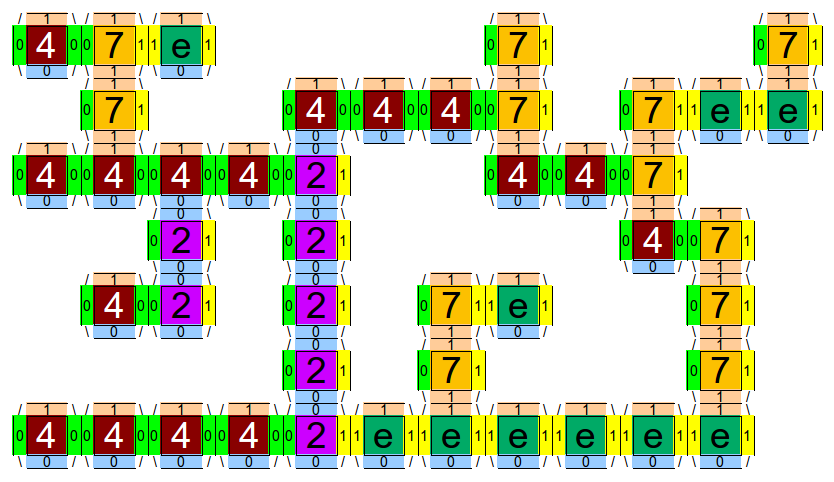}\hspace{-.2cm}
\includegraphics[scale=0.07]{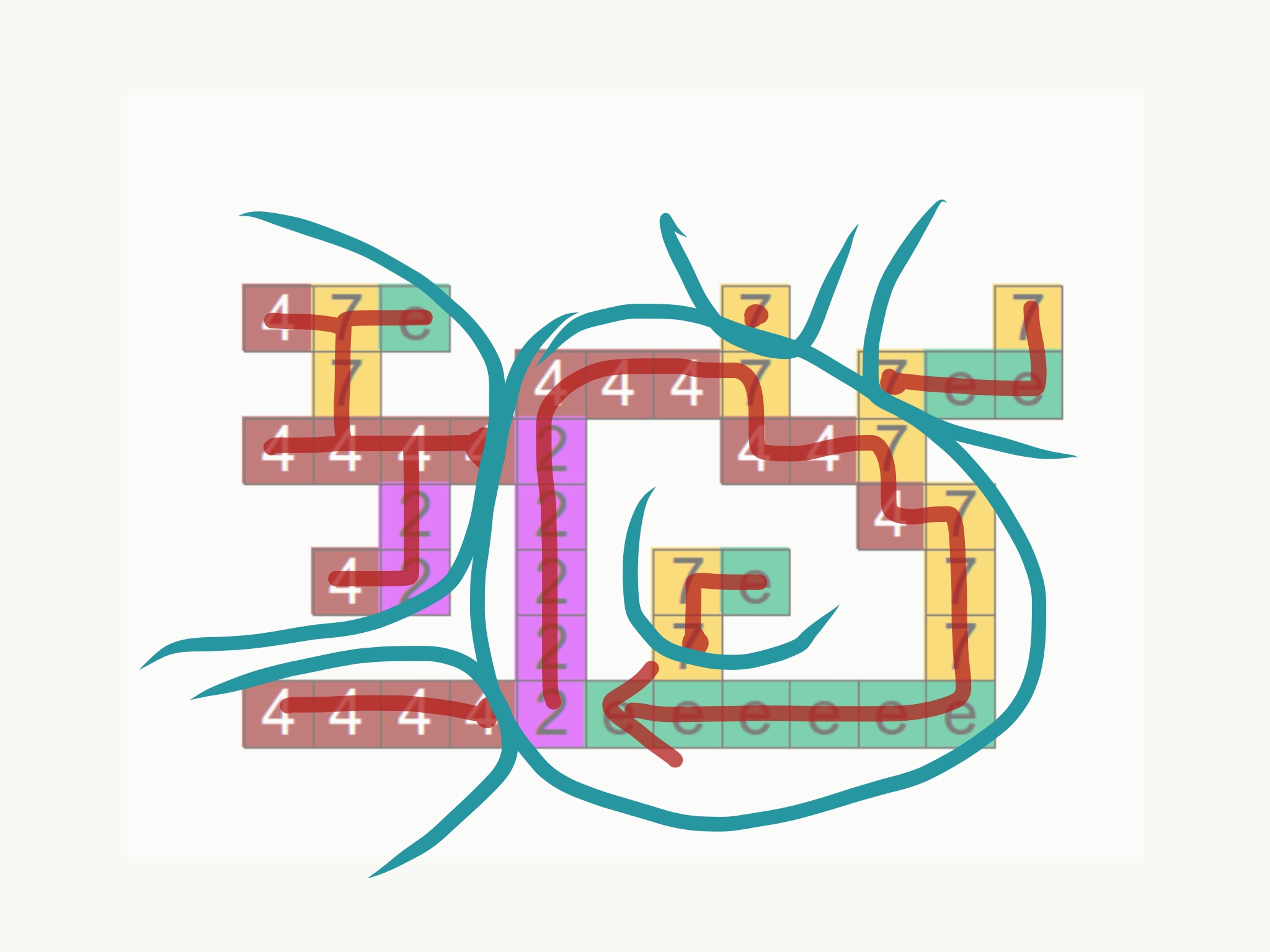}}\vspace{-.3cm}
\caption{A 2D word and an accepting tiling}\label{nf111}\vspace{-.5cm}
\end{figure}

The 2D automata, used to define the structure of VO in this paper, are specified with subsets of the tiles below and have as labels (or colors) for external borders, in accepted words, those shown in the figure: 
\vsp\\
\centerline{\includegraphics[scale=0.4]{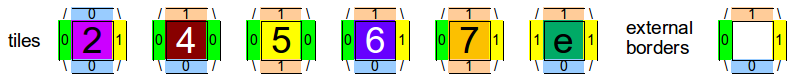}}
\vsp\\
This means, we look for tilings where: (1) any two connected tiles agree on the common side; and (2) any external west / north / east / south border has label 0 (green) / 1 (red) / 1 (yellow) / 0 (blue), respectively.   

Actually, for this tiles, the border labels are obtained placing the digits of the 4-digits binary representation of the body label on the west/north/east/south borders (in this order); the same convention is used for specifying the accepting external borders. For instance, 6 has the binary representation 0110, so its labels are 0/1/1/0 on the west/north/east/south borders. This automaton is shortly denoted by $24567e\_6$. Fig.~\ref{nf111} shows an accepted tiling and the associated word.

Sometimes we use an alternative notation for the tiles 2,4,5,6,7,e as
\vsp\\
\centerline{\includegraphics[scale=0.3]{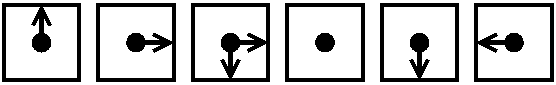}}
\nvsp\\ 
respectively, where each small arrow indicates a direction where we have to continue the tiling process to meet the external label goal, here specified by `6'. With this representation, we see a clockwise cycle in Fig.~\ref{nf111} going, from the bottom row, up with 2's, to right with 4's, down with 7, to right with 4's, down with 7, to right with 4, down with 7's, and to left with e's.

\begin{figure}
\centerline{\includegraphics[scale=0.2]{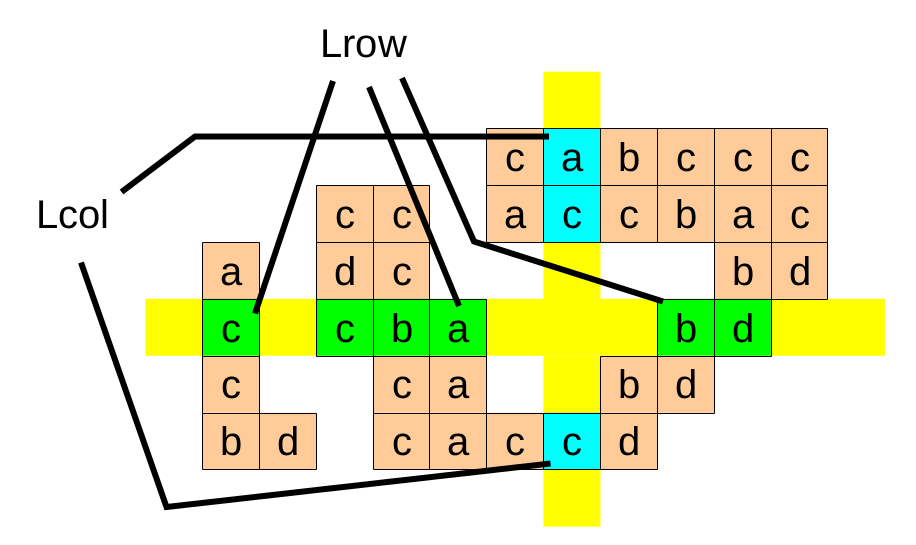}}
\caption{2D languages as the product of two 1D languages (one $Lrow$ is for rows, the other $Lcol$ for columns)}\label{f-lpp}
\end{figure}

\paragraph{$\otimes$-product} An equivalent\foo{The equivalence is valid under mild conditions, e.g., provided tiles have different body labels. Indeed, given a 2D automaton $T$ and its projections $T_r$ (on rows) and $T_c$ (on columns) an accepted scenario for $T$ has rows and columns accepted by $T_r$ and $T_c$, respectively and conversely. As we mentioned before, for arbitrary 2D automata, one has to add renaming to get equivalent $\otimes$-product specifications.} definition of a regular 2D-languages is as the $\otimes$-product of two regular languages, one for rows and the other for columns. By definition, a 2D word is in the $\otimes$-product $R\otimes C$ if all its maximal 1D words on rows are in $R$ and all its maximal 1D words on columns are in $C$. The $\otimes$-product operation is illustrated in Fig.~\ref{f-lpp}.  

\out{
\begin{figure*}
\begin{tabular}{c@{\hsp}c@{\hsp}c@{\hsp}c@{\hsp}c}
\includegraphics[scale=0.2]{fig/247e-6_w1} 
& \includegraphics[scale=0.12]{fig/247e-6_s1}
& \includegraphics[scale=0.3]{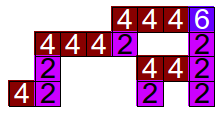}
& \includegraphics[scale=0.3]{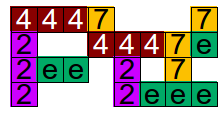}
& \includegraphics[scale=0.3]{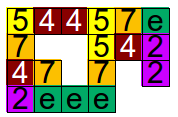}\\
(a) & (b) & (c) & (d) & (e)
\end{tabular}
\caption{A 2D word (a) and an accepting tiling (b); then examples of words in the languages $L(246\_6)$ (c), $L(247e\_6)$ (d), and $L(247e5\_6)$ (e)}\label{nf1}
\end{figure*}
}

\paragraph{2D regular expressions}
Regular\foo{The term ``regular'' is used in a loose sense here and should not be interpreted as an equivalent way of defining regular 2D-languages. Indeed, it is an open question whether there is a particular class of 2D expressions, equivalent to 2D automata, defined above.}  2D-expressions, for arbitrary shape words, are more difficult to define. Roughly speaking, they  specify mechanisms to generate words using powerful composition operations, based on restrictions on the shared border of composing words in the composed words. These composition operators are used in combination with recursive definitions to specify classes of 2D-languages. A few nontrivial examples are  presented in Appendix B (and in  \cite{DBLP:conf/RelMiCS/Stefanescu15}). 

For a simple example, consider the tree structure on the right, defined by $4^*(2+6)\otimes (4+6)2^*$. This language can be generated, with regular 2d expressions, as follows:
\vspace{.5cm}\\${}$\hfill{\includegraphics[scale=0.4]{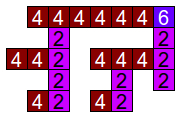}}\vspace{-3.cm}\\
\begin{verbatim}
X1 = 2 + X1 (n=s) 2 -- vertical bars of 2’s;
X2 = 4 + X2 (e=w) 4 -- horizontal bars of 4’s;
X3 = 6 + X1 (n<s) X3 + X2 (e<w) X3 -- an incorrect, first attempt
X4 = 6 + X1 ((n<s) & !(s#n) & !(e#w) & !(w#e) X4 -- correct version
       + X2 ((e<w) & !(w#e) & !(n#s) & !(s#n) X4
\end{verbatim}
Intuitively, X1 generates columns of 2-cells, starting with a 2-cell and recursively connecting 2-cells at its top (``\texttt{X1 (n=s) 2}'' means: connect X1 and a 2-cell such that the \textit{n}orth border of X1 is equal to the \textit{s}outh border of this 2-cell). Similarly, X2 generates rows of 4-cells (recursively connecting to their \textit{e}ast border to the \textit{w}est border of a 4-cell). A first attempt, specified by X3, to generate trees is to start with a 6-cell and recursively add columns of 2-cells (with their north border included in the south border of the already obtained word; the restriction ``\texttt{X1 (n<s) X3}'' says the north border of X1 is included in the south border of X3), and rows of 4-cells (with their east border included in the west border of the already obtained word). This is slightly incorrect, as, for instance, two colums of 2-cells may connect each-other. To avoid this, the extra conditions, seen in X4, are added; here, a condition as  \verb+X !(s#n) Y+ is read ``it is not true (the `!' symbol) that the south border of X and the north border of Y share a (nonempty) common border in the composed word (the `\#' symbol denotes non-empty intersection)''.

\paragraph{Example}
\emph{Rings with Attached Trees}: This language, denoted $RAT$, consists of 2D words having the following structure: there is an attractor ring, either going in a clockwise direction  or in an anticlockwise direction, and trees attached on this ring. An example is presented in Fig.~\ref{nf111}.
\be
\item A particular 2D automaton for RAT is specified by the tiles \{2,4,7,e\} and the border condition 0/1/1/0 for the west/nort/east/south borders, shortly denoted by $RAT = L(237e\_6)$. 
\item The $RAT$ language may equivalently be defined as the product of two regular languages $4^*(2+7)e^*$ (for rows) and $7^*(4+e)2^*$ (for columns), shortly written $RAT=4^*(2+7)e^*\otimes 7^*(4+e)2^*$.
\item A regular expression for this language may be found in the Appendix B.
\ee
The first two representations of this language are short, but it is relatively difficult to understand which 2D-words are specified. The last representation is relatively long, but it is easier to understand which 2D-words are generated. In other words, specifications like those in (2) are more like a statement of a problem (i.e., find which 2D-words have the specified forms on rows and columns), while equivalent presentations as those in (3) are solutions to those problems. Moreover, the specifications with regular 2D-expressions are compositional, and may be used to get compositional techniques for aggregating virtual organisms.

\subsection{Examples of regular 2D patterns}

\paragraph{Trees} 
$$Tr=L(246\_6)=4^*(2+6)\otimes (4+6)2^*$$
This pattern describes trees, with: (1) a 6-cell in the top-right corner; (2) horizontal edges of 4's, connected to the tree with their east border; (3) and vertical edges of 2's, connected to the tree with their north border. An example is below, represented in both ways.
\vsp\\
\centerline{\includegraphics[scale=0.3]{fig/246-6_w1}
\hspace{2cm}\includegraphics[scale=0.4]{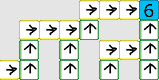}}
\vsp

\paragraph{Rings with Attached Trees}
$$RAT=L(247e\_6)=4^*(2+7)e^*\otimes 7^*(4+e)2^*$$ 
This, already mentioned pattern, consists of:\bi
\item a ring, containing a clockwise or a counter-clockwise cycle of cells 2,4,7,e;
\item trees with horizontal edges of 4's or of e's, vertical edges of 2's or of 7's, and with their roots fixed on the ring (these trees may be either in the interior or the exterior areas separated by the ring).
\ei
An example, represented in both ways, is\foo{A larger example is shown in Fig.~\ref{nf111}}:\svsp\\
\centerline{\includegraphics[scale=0.3]{fig/247e-6_w2}
\hsp\includegraphics[scale=0.4]{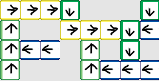}}

\paragraph{Connected Rings with Attached Trees} $$CRAT
=L(2457e\_6)=(4+5)^*(2+7)e^*\otimes (7+5)^*(4+e)2^*$$
The description of this language is in 2 steps:\bi
\item first, the pattern $L(56\_6)$ is analyzed;
\item then, the 6's cells in these configurations are replaces by configurations described by the pattern $L(247e\_6)$.
\ei
The first language, described by the product $5^*6\otimes 5^*6$, consists of 6-cells, placed along the second diagonal, with a stair-like shape in the top-left area filled with 5-cells. Then, roughly speaking, the full language results by replacing the 6-cells in these pattern with RAT words. Briefly, the 5-cells act as binders, linking leafs of RAT words in the particular shape indicated by the $L(56\_6)$ pattern. 

A simple example is:\foo{The meaning of the double-arrow cell, replacing 5-cell, should be clear: one has to continue the tiling in those two directions to meet the external border restriction.} \vsp\\
\centerline{\includegraphics[scale=0.3]{fig/2457e-6_w1}\hspace{1cm}
\hsp\includegraphics[scale=0.4]{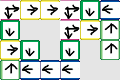}}
\vsp\\
A more complex example is below, together with a decomposition emphasizing the role of the 5-cells to connect words in $L(247e\_6)$ (i.e., rings with attached trees).\vsp\\
\centerline{\includegraphics[scale=0.2]{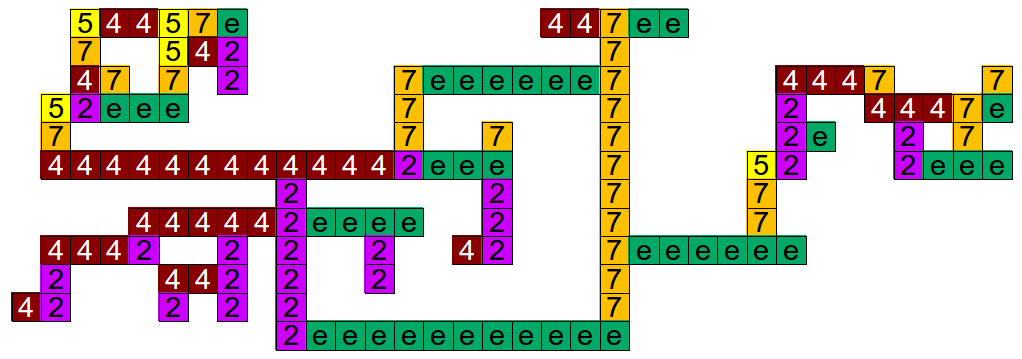}\hsp
\includegraphics[scale=0.2]{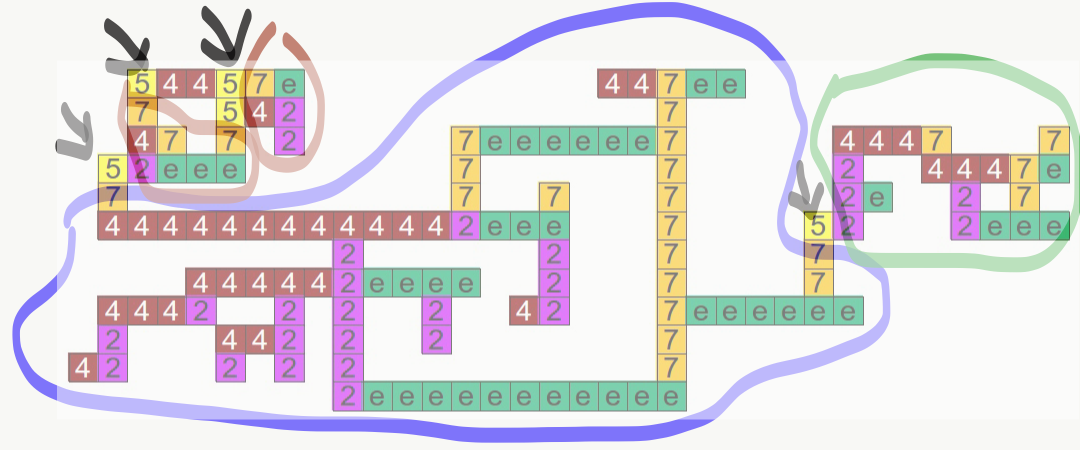}}
\vsp

\section{Virtual organisms}

In this section we describe virtual organisms obtained by adding functionalities to regular 2D structures. The cells in a 2D word are interpreted as nodes in a network defined by the word structure and they are able to perform local computations and communications with the neighboring nodes. 

\subsection{What is a virtual organisms?}
Informally, a \emph{virtual organism} is defined by a structure and a functionality mapped on that structure. The structure is specified by a word in a (regular) 2D pattern; this structure in further interpreted as a network of nodes performing local computations and communications with the neighbouring nodes. Then, a collection of overlapping functions are implemented on top of this network. For a fixed structure (i.e., no reconfiguration), the overall behavior of the network is specified by an \emph{input-output stream processing relation} between the streams on the external border interfaces through which the organism communicates with its external environment. 

In the simplest case, the specification of a virtual organism behavior is as the overlapping of multiple independent functions. Each function is defined under the hypothesis that the network is running that function, only (noninterference). Then, the overlapping of the functions in the network have to assure preservation of independent functions' behavior. This latter condition may be asked in a strict form (i.e., the independent stream processing relations are preserved), or in a more relaxed form, namely up to processing delays in the external streams -- the latter in especially the case when the nodes' resources are limited and delays in the precessing of some of them is inevitable.

In other cases, certain functions may be introduced on top of other functions, hence a smaller or a larger amount of interference between functions may occur. For instance, a function using position-aware node functionality may depend on another function detecting and broadcasting the organism current structure.

A more complicate case is when the structure itself is dynamically changed by reconfiguration. In that cases, even the external interface of the organism may change, hence the definition of the input-output stream processing relation is more involved: during the reconfiguration steps, the impact of reconfiguration on the external streams has to be specified, as well.

\paragraph{Remark} An important point, to be clarified here, is that communication in a VO is not limited by the topology of connected nodes. The given structures, specified with 2D words, are needed for coordination of nodes to support the VOs basic functionalities. For instance, if a structure as in Fig.~\ref{nf111} is used for a VO, than it is possible that, in a particular deployment, all virtual nodes in an attached tree are handed by the same physical node, making more communications possible. The situation is similar to that found in ring distributed termination protocol \cite{dijkstra1986derivation}, or in the CHORD protocol \cite{stoica2001chord}, where the nodes are placed in a ring, but for additional functionalities (e.g., exchanging jobs, or records) more communication links may be used.

\subsection{Examples of virtual organisms}

Before briefly presenting a few simple virtual organism, we describe a common, useful functionality: detecting and broadcasting the organism structure. This function is triggered at the beginning of the computation and after each reconfiguration step. Its  implementation may be either a general one (designed for all structures), or a specific, more efficient variant, exploiting the current particular structure.

\paragraph{TreeCollector (TC-) organisms}
These TC-organisms are developed on top of the $Tr$ pattern. Recall that a $Tr$-word is a tree with the root, labeled by 6, placed in the top-right corner, with horizontal rows of 4's, and vertical columns of 2's.

Except for general structure detection and reconfiguration functionalities, a TC-organism have the following specific functionality\foo{As a distinction occurs between the behavior of a leaf node and a non-leaf node, it is clear that this functionality depends on the structure detection functionality.}:
\bi\item it collects, at the root, a flow of items captured via the tree leafs.\ei 
The scenario we consider assume several sources are present in the TC-organism's area and the amount of flow (or volume) each leaf can capture depends on its distance to the sources. 

\paragraph{FeedingCell (FC-) organisms}
These FC-organisms use the $RAT$ pattern. Recall that a RAT-word consists of a ring of nodes 2,4,7,e (ordered in a specific clockwise or anticlockwise cycle), with attached trees on both sides. 

The specific functionality of a FC-organism is to 
\bi\item collect items from sources, placed in the external area of the ring, via the external trees, distribute them along the ring, and release them via the internal trees to reach certain target points in the internal area.\ei

\paragraph{ConnectedFeedingCells (CFC-) organisms}
CFC-organisms are build up on top of the more complicate $CRAT$ pattern. Roughly speaking, a CRAT-word consists of a collection of RAT-words, connected via 5's nodes.

An illustration of a specific functionality, to be added to the FC-organism's functionalities, may be to 
\bi\item reverse the flow in a certain FC-organism component, allowing to take items from its internal area, and pass the items, via trees and the ring, to the external tree leafs, then use them, via 5's nodes, to feed other FC-organisms included in the CFC-organism.\ei

\subsection{Reconfiguration of virtual organisms}

Virtual organisms may adapt to the changes in the environment modifying their structure and/or functionality. In this paper we focus on the first possibility, by allowing virtual organisms to reconfigure themselves, replacing their structure by another structure in the same 2D pattern.

There are two different types of structure reconfiguration:\be
\item \emph{conservative reconfiguration}, preserving the cells
\item \emph{elastic reconfiguration}, allowing to add or remove cells.
\ee

\begin{figure*}
\includegraphics[scale=0.35]{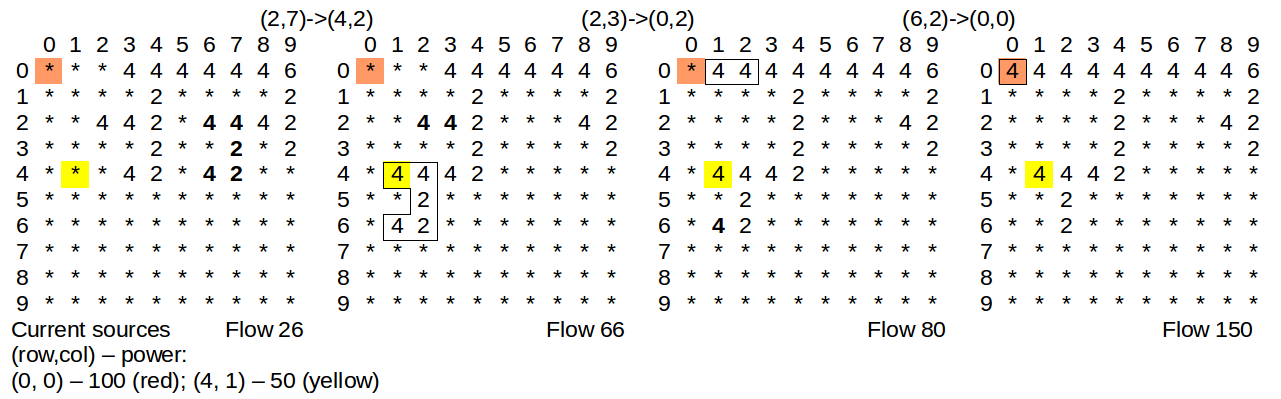}
\caption{Reconfiguration of TreeCollector organisms}\label{f77}
\end{figure*}

An example of a chain of conservative reconfigurations, with reference to our simulations described in the next section, is presented in Fig.~\ref{f77}, using TC-organisms. 
The flow captured by cell $(0,3)$ from source $(4,1)$, as described by formula (1) below, is $50/(manhattandist + 1)^2 = 50/(2+4+1)^2$, while from source $(0,0)$ is $100/(3+1)^2$. Summing up for all leaf nodes and sources leads to a total flow of $26.83$, rounded to $26$ in Fig.~\ref{f77}. 
The subtree part in boldface in the first configuration is moving its root position from (2,7) to (4,2), increasing the flow from 26 to 66. Two more reconfigurations (moving the node (2,3) to (0,2), then the node (6,1) to (0,0)) lead to an optimal structure, increasing the flow to 150.

In the next section, we also describe an elastic reconfiguration model for TC-organisms, based on the cost function described by formula (2). To optimize such a cost function, during a reconfiguration step new nodes may be rented or old nodes may be released.

FC-organisms are more complex and they allow for a broader class of reconfigurations. As a conservative reconfiguration, we mention the possibility to detach a tree from the external part of the ring and attach it to the internal part of the ring. In the elastic reconfiguration case, the ring and the tree components may also change their shape or dimension. 

\subsection{Compositionality}

A very important feature of the model sketched above is \emph{compositionality}. While not explicitly stated, it is intuitively clear that the later introduced and more complex organisms, presented in a previous subsection, may be obtained by appropriate compositions from simpler organisms. For instance, FC-organisms may be specified as a composite of TC-organisms with a ring-passing organism (circularly passing items in the ring). Similarly, CFC-organisms result from composing FC-organisms, glued with 5's cells.

The virtual organisms' composition mechanisms are based on composition mechanisms used for regular 2D languages, illustrated in Appendix B (or, in extenso, in \cite{DBLP:conf/RelMiCS/Stefanescu15}).

\subsection{A detailed example: The TreeCollector organism}

\newcommand{\cdoi}{\includegraphics[scale=0.2]{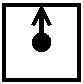}}
\newcommand{\cpatru}{\includegraphics[scale=0.2]{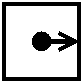}}
\newcommand{\csase}{\includegraphics[scale=0.2]{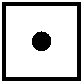}}

The structure of the TreeCollector organism (TC-organism) is specified by a word in the language $Tr=4^*(2+6) \otimes (4+6)2^*$. A word in $Tr$ represents a tree, with the tree root, labeled by `6', placed in the top-right corner, with (horizontal) rows $4^*6$ or $4^*2$, and (vertical) columns $42^*$ or $62^*$. (Notice that no two $4$-cells are placed in the same column, one on top of the other, or two $2$-cells are placed in the some row, one near the other, hence a valid tree structure is specified.) 

\subsubsection{Leaf-to-root item collection:}
The basic functionality of the TC-organism is to collect items via its leaf cells, then to pass them, from cell to cell, till the collected items arrive to the root cell. Each cell has a local knowledge on the VO structure, knowing its neighbors at the west/north/east/south borders, i.e., whether there is a neighbor on a specific side and, if it is, what is the data type for communicating with that neighboring cell.

\begin{figure}
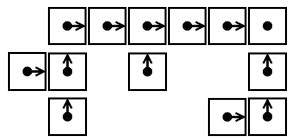

\centerline{
\begin{tabular}{l@{}l@{}l@{}l@{}l@{}l@{}l}
&\cpatru&\cpatru&\cpatru&\cpatru&\cpatru&\csase\vspace{-1mm}\\
\cpatru&\cdoi&&\cdoi&&&\cdoi\vspace{-1mm}\\
&\cdoi&&&&\cpatru&\cdoi
\end{tabular}}
\caption{Collecting flow in TC-organism}\label{f-tc1}
\end{figure}

The flow direction for the TC-organism collection operation is illustrated in Fig.~\ref{f-tc1}. The code for the basic collection step of a $2$-cell is bellow
\begin{verbatim}
Code1 of a 2-cell, for items collection functionality:
  if (leaf) {flow = collect();} else {flow = 0;};
  if (WestNeighbor != nil) {recv(f1,WestNeighbor); flow += f1;};
  if (SouthNeighbor != nil) {recv(f2,SouthNeighbor); flow += f2;};
  send(flow,NorthNeighbor);	
\end{verbatim}
The code for a $4$-cell is similar, but the last statement is changed to send the collected flow to the eastern neighbor, i.e., \verb+ send(flow,EastNeighbor)+. The code for the $6$-cell is obtained deleting the last statement.

\subsubsection{Structure detection:}
With a specific data structure for items, the collection function may be used to obtain the actual structure of a TC-organism. For instance, use a function $Tree(nod, tree1, tree2)$ to construct a tree with the root $node$, with an horizontal branch $tree1$ and a vertical branch $tree2$, where one or both $tree1$, $tree2$ may be $nil$. This process is triggered by the root cell, sending a message down to the tree to start this structure-identification process.

The implementation is: 
\begin{verbatim}
Code2 of a 2-cell, for structure-detection functionality:
  if (WestNeighbor != nil) {recv(t1,WestNeighbor);
    } else {t1 = nil;};
  if (SouthNeighbor != nil) {recv(t2,SouthNeighbor);
    } else {t1 = nil;};
  send(Tree(myId,t1,t2),NorthNeighbor);	
\end{verbatim}
The code for a \texttt{4-cell} is similar, but, in the last line, the send is to the \texttt{EastNeighbor}, while the code for the \texttt{6-cell} is obtained deleting the last line. 

\subsubsection{Reconfiguration:}

The implementation of reconfiguration for TreeCollector organisms is relatively easy: 
\begin{quote}
Suppose the root cell `$6$', analyzing the received items' flow, decides to make a structure reconfiguration, for instance, to increase the collected flow. It will send a message down to the tree to change the local neighboring topology to the selected new one. For example, to move a subtree, with root $X$, having label `$4$', to be linked as a child of a node $Y$ of the tree, the following changes are needed: the west link of the parent of $X$ becomes $nil$; the east link of $X$ becomes $Y$; and the west link of $Y$ is changed from $nil$ to $X$. The case for label `$2$' is similar.
\end{quote}

\subsubsection{The overall behavior}
The overall behavior of a TC-organism is specified by a schedule of the functionalities supported by the organism. A typical one may be to repeat the collection process, until the root triggers the reconfiguration process, then to repeat these steps. A variation may be, for instance, in the elastic reconfiguration case, if the new cells are locally attached to the structure; then, before triggering the reconfiguration process, the root has to invoke the structure-detection functionality, to know the actual tree structure. Other variations may take into account the communication delay of passing the items along the tree, the bandwidth of communication, etc.

\section{Simulation of virtual organisms behaviours}

To test the theoretical aspects discussed in this paper, we created an open source framework, publicly available at github link 
\vsp\\\centerline{https://github.com/AdaptiveOrganisms/Simulator}\vsp\\ 
It can be executed in a distributed environment and, by using functional hooks, can be reused for various use-cases. The framework's folder also contains a visual interactive simulation and a movie that shows visually the average flow difference between a static TC-organism and an adaptive one.

\subsection{Simulation framework}\label{subs:reconfig}

Our general case for simulation is an environment where there exists a dynamic set of data sources named $Sources$, each one having a variable data rate in time $Source_{i}.power$. This set can be modified at any point in time, new sources can be added, removed or have their data rate modified at user intervention or according to a random simulation.  A binary tree language, like $Tr$, used for TC-organisms, can represent a connection of resources that capture data from sources and move them up to the root of the tree. Only the leafs of the tree can capture data, while the internal nodes' role is to transport it to the root. One concrete usage of this representation could be a wireless network, where sources are requests emitters, and the role of the network is to capture as much as possible data and transmit it to the root, which can, in turn, represent a processing endpoint for these requests. 

As shown in the evaluation section, since $Sources$ set is dynamic, reconfigurations in the tree are needed to keep a good dataflow to the root. These reconfigurations are based on cutting a subtree rooted in any of the nodes and pasting it to another position, provided it still represents a word in $Tr$. The formula used to determine the maximum that a leaf node can capture per unit of time is given by formula (\ref{eq:capture}), namely leaf nodes can capture the data flow emitted by any source, proportionally to the squared Manhattan distance between them, increased by 1. A source capacity is considered limited and it serves the leaf nodes up to its maximum data flow per unit of time. If the leaf nodes around source can capture more data flow that the source can handle, the amount each leaf node gets is  proportional to the normalized to the source, comparing to other competing leaf nodes.\nvsp

\begin{equation} \label{eq:capture}
Capture(leaf) = \sum_{s \in Sources} \frac{s.power}{dist(leaf, s)^2}
\end{equation}\nvsp

From a greedy perspective, it makes sense to move parts of the tree as close as possible to the data sources. However, each node in the tree is considered to have a maximum (predefined) data flow per unit of time specified by user, $F_{max}$, which complicates a greedy solution attempt. Our target is to simulate situations where nodes are considered processes, and limiting the maximum dataflow on each node per unit of time can represent either the throughput of the node or the maximum communication bandwidth between nodes.

The regular expressions used to specify constraints between the nodes can have various significations in practice. For instance, in the IoT the  constraints defined by the regular expression can signify the pattern of the accepted resources that can be connected together to transport certain data flows. In practice, these restrictions can stem either from hardware limitations or security concerns.

\subsection{Implementation details}

In this subsection, we discuss only the high-level aspects of the implementation. Users interested in more details about the implementation and pseudocode of the simulator are advised to read the documentation in our github link. 

The implementation of the framework is using message passing mechanisms at its core. Each node can be represented by a different process instance (we name it $Cell$). The type of the process instance is decided on the respective's node symbol. The links in the tree represent dataflow connections between pairs or processes, from children to their parents. The end of this dataflow is the process instantiated for the root of the tree. The role of each $Cell$ is to try to capture as much as possible data flow it can from sources, respecting the maximum flow per unit of time constraints.

Reconfiguration means moving the location of processes and the links between them in a way that optimizes the overall data flow per unit of time. It can be requested by user manually or automatically triggered based on some conditions (e.g., sources configuration have changed since the last check). Intuitively, processes must physically move closer to sources. However, because of the maximum flow constraint and the dynamicity of sources set, an algorithm that finds the optimal nodes that should be moved and their new position, while still holding the row and column languages constraints, is not trivial to find. Each $Cell$ process, excepting the root, computes the best option that it can offer to the overall data flow per unit of time by cutting its entire subtree and pasting at all available positions that match the constraints in its local view. In the current implementation, the search for the optimal tree modifications that matches new situations is a brute force search that tries every possible position for pasting. Each option is tested against a data flow evaluation function that tries to estimate the data flow of the network in case that the selected operation is being done. These local best results are gathered by the root of the tree which in turn select the option that gets the maximum flow at that point. 

The elastic model can be enabled optionally and runs a process after each reconfiguration step that decides if resources can be rented or given back (the restriction is to give back only resources that were rented). The purpose of this is to simulate a basic operation in cloud computing were resources can be rented at runtime to solve workload issues in a system. At a lower level, the elastic model process tries first to insert a new node in the tree either adjacent to a leaf (considering the appropriate left or down side of the leaf), or in the place of an already existing resources and   moving the previous entire subtree either left or down. The user can configure the benefit per each unit of flow, the count limit and cost of each type of resource that can be rented. Equation~(\ref{eq:Cost}) gives the cost at any point in time considering the rented resources, considering that the current average flow in the observed time frame is $AvgFlow$, that the benefit per each unit of flow is set by variable $BenefitUnitOfFlow$, and the set of rented nodes is $Rented$. Going back to the purpose of the elastic model, its role is to maximize this function at runtime.
\begin{equation}\label{eq:Cost}
\begin{array}{rcl}
\hspace*{-.3cm}Cost(Tree) &=& AvgFlow * BenefitUnitOfFlow\\
&& - \sum_{S \in Rented} Cost(S)
\end{array}
\end{equation}

\subsection{Evaluation}

First, we want to check the benefits of the reconfiguration operation, i.e., how close to the optimal tree solution we can get by applying it successively. Then, we compare the flow in the initial static optimal tree and the dynamic version, considering the cost of reconfiguration and uniform distribution of add/remove/modify events inside $Sources$ set. Finally, we analyze the elastic model and how much can it improve the benefit.

\subsubsection{Reconfiguration evaluation}

Since finding the optimal tree considering a given set of Sources, the $Tr$ regular pattern and a specified number of cells (resources) of each type is hard and because the potential number of binary trees can grow exponentially (Catalan numbers), we decided to apply an inverse Monte Carlo method for finding a tree which is as close to optimal as possible, within a given time limit. Thus, a random tree was generated and fixed, then the $Sources$ was randomly sampled. For each new sample, the initial random tree was also reconfigured until its  maximum flow couldn't be improved anymore. The configuration of $Sources$ and reconfigured initial random tree that got the maximum flow was considered as ``optimal''. Then, the simulation randomizes the optimal tree structure by keeping the same numbers of different symbols in the tree. We are interested to see how close does the randomized tree's maximum flow can get after several reconfigurations compared to the optimal flow. 

 By averaging the results for many attempts like the one described,  on trees with 20-25 nodes and uniform distribution of symbols, the ratio obtained between the flow after reconfigurations and the optimal flow was, in average, $0.96$. This was achieved after an average $3.69$ reconfiguration steps, which suggest that reconfiguration can be a fast method for achieving a good flow starting with a random tree.

\subsubsection{Dynamic versus static tree structure}
For this evaluation, we considered the average flow gathered for a number of scenarios, that started with an optimal tree and sources configurations, and compared it against a random initial tree with the same symbols count (resources) and which was allowed to reconfigure. The number of ticks used in the simulation for each scenario was a variable $N$ ($=100$, in our case) and the probability of having a source modification event (add/remove/modify) within the existing set at each tick was represented by another variable $P$ (considered to be $0.2$).
As mentioned in \ref{subs:reconfig}, only the root node of the subtree that moves during reconfiguration affects the flow of the overall tree. For a given number of ticks, it can't send further the data flow gathered by its subtree. As shown in Table \ref{table:flowpercents}, the time needed for the root to attach to a new parent affects how well does the dynamic structure behaves.

\begin{table}
\centerline{\begin{tabular} { | p{6.5cm} | p{6.5cm} |} 
\hline Percent (in ``ticks'') of a single reconfiguration cost  from the total scenario time & Average improvement of flow in dynamic versus static tree \\ 
\hline 1\% & 85\% \\ \hline 5\% & 27\% \\ \hline 10\% & 11\% \\  \hline
\end{tabular}}\vspace{.3cm}
\caption{Tree structure with reconfiguration vs. optimal static tree}
\label{table:flowpercents}
\end{table}

The results can also depend on variable $F_{max}$, and the data rate of each source. In our simulation, $F_{max}$ was 10000 and each source of the three sources (on average) used in the simulation had a data  rate between 100-1000. An observation is that at the beginning of each scenario, the random tree had an worser data flow, but by using reconfigurations it fastly got to the performance of the optimal tree. However, after a few reconfiguration steps and by considering the dynamicity of $Sources$ set, the dynamic tree got better in terms of maximum flow per tick.  We conclude that overall, the reconfiguration operation is usefully for increasing the maximum flow if the sources set is dynamic.

\subsubsection{Elastic model enabled}
To evaluate this model we created a set of test samples with the same number of nodes as before (20-25) and using up to 8-10 number of additional resource limit that can be rented at runtime. To observe if the tree simulation tries to rent the resources as much as possible without having their cost influencing the decision, we set as benefit for each unit of flow to be $10$ times higher than the cost of each individual resource. In this situation, the algorithm added resources in different places where the flow bottlenecks existed because of the maximum data flow constraint per unit of time on each node. The average flow (and benefit) increased with  $~55\%$. The model also showed optimal results when giving back unused resources. 

However, more work on this elastic strategy should be done since we considered that rented resources are available to use instantaneously, which is usually not the case and some time is needed to configure the new resources. For this new requirement to work correctly, we could employ some machine learning mechanisms that are able to predict the patterns in a system and know before when to start renting or giving up resources to make sure that the peek costs are achieved.
 
\subsection{Implementing virtual organisms in AGAPIA}

Agapia is a DSL structured parallel programming language for interactive systems (see \cite{dr-st08a,pss07,DBLP:conf/ispdc/Paduraru14,p15}, or the recent survey \cite{ecsa}).  
The current version of its compiler, allowing to quickly design and run HPC applications on MPI and OpenMP platforms, is open source with online deployment:
\vsp\\\centerline{https://github.com/AGAPIA/CompilerAndTools}\vsp\\ 
The package contains a manual on how to use it, and sample problems implemented and ready to run.
The starting point for building Agapia programs are interactive modules. Modules have data on interfaces, but no labels to allow for control and interaction coordination patterns. Traditionally, a program is seen as a transformation with input data on the west and north interfaces and output data at the east and south interfaces.

While AGAPIA promotes a loosely-coupled architecture, it is also able to impose restrictions over the communicating components in distributed systems. More, in AGAPIA, the default communication model is between sender and receiver directly not through a centralized node (such as a broker node), which resolves the message overloading problem in the publish-subscriber paradigm. However, inside programs users are free to mix between the two paradigms.
In applications composed of several code pieces it is important nowadays to provide ways to parallelize the execution process. AGAPIA provides a background parallel execution of programs without needing user to write explicit parallel work creation or synchronization. The runtime component of AGAPIA tries to maximize the parallelism inside an application by mapping ready for execution instances of programs (i.e. instances with no input dependencies) to workers: threads - when being executed in the shared memory model, or processes on different machines - when executed in the distributed memory model.

Regarding performance, as evaluated in \cite{p15}, the only overhead of the AGAPIA language compared to a hand-written application in a low-level programming language comes from communicating data between modules when performing interpretation. This internal overhead was evaluated at less than $7\%$ for the shared memory model and around $15\%$ for the distributed execution model. It is important to state that the C/C++ code inside modules is still compiled and added as native code in the binary file. The only interpreted part of an application written in AGAPIA are its high-level statements and composition operators that usually drive the execution of the native code. Overall, the implementations in AGAPIA can prove to be faster to write, maintain and extend, less error prone at a small performance incur.

Compared with the virtual organism framework, the Agapia model use 2D regular patterns, with 1D used for space and 1D for time. The natural passing from VO simulations to Agapia programs is to flatten the 2D virtual organism spatial structure into 1D. Then, most of the simulation is obtained for free form: (1) cells code; and (2) some high-level description of the schedule mechanism used for organism functionalities. 

The current users' requirements estimations for a language that supports virtual organisms could be defined as:
\be\item
Specification of a virtual organism structure as input;
\item Allow users to inject custom functionality for its cells.
\item Deploy it on a distributed system to support different applications (e.g. such as the ones in the IoT domain).
\ee
More details are provided into Appendix~\ref{ap-a}.

\section{Related and future work}

The virtual organism model is simple, compositional, expressive and naturally incorporate self-adaptive systems requirements. It has roots in two well-known popular approaches (i.e., 1D regular languages \cite{kleene1951representation, conway2012regular} and dataflow networks model \cite{dennis1974first,halbwachs1991synchronous}), hence it is expected to be easily adopted. For instance, the programs written in TensorFlow \cite{abadi2016tensorflow} produce virtual organisms (usually not reconfigurable). Here, we briefly present the directions for our planned future work and connections to related work.   

\paragraph{Future work}

Our planned future work, for virtual organisms, is currently split into four categories:
\begin{enumerate}
\item A library of VOs templates which user can use out of the box, or modify existing policies to test desired functionality.
\item Programming language design to allow specification of VOs functionality.
\item Automatic reconfiguration strategies using Machine Learning techniques.
\item Creation of competitive environments for testing reconfiguration strategies.
\end{enumerate}

The first item is probably the most important for model adoption. We plan to start by selecting a field (e.g., one from those mentioned in the introduction, namely IoT, self-adaptive software, mobile networking, robotics, etc.), to identify and analyze a class of relevant distributed applications from that field, then to decompose those applications looking for key parts which deserve to be fully developed as VOs. Hopefully, the identified VOs may be good enough to produce composed VOs replacing those applications and subsequently be used for easy development of new applications.

The plan for the second item is to improve either AGAPIA support or build another language to specify VOs. The main purposes of this language would be: (1) Flexibile VO specification and deployment; (2) Users should be able to overwrite the policy of each node type in a VO; (3) Elastic limitation and behaviors of rent and lease of resources should be specified; (4) Allow easy interaction with a VO at runtime, i.e., change policies by updating a handler function in the source code specification of the model (such specifications could include scripted user behaviors for reconfigurations, getting observations from the environment and letting a machine learning model to infer these and get a decision, or real-time training of a model).

For the third item, the purpose is first to learn patterns from real data, where a VO problem was instanced on, and improve the reconfiguration decision to get improved throughput. The aim is to predict the reconfiguration in advance, such that the system is able to rent or release new resources, or construct parts of the organism in safer steps. Recurrent Neural Networks can be a good fit for the models since they could use a memory of previous actions taken and environment state, to take the next decision. Also, we plan to use Deep Reinforcement Learning to improve the model after initial fitting from real data, where actions are typically represented by reconfigurations, observations could be environment states and predictions of sources, while rewards can be related to the delta of the captured flow between different moments of time. For evaluation purposes, we plan to implement these on some classes of the TreeCollector problem considering either static or mobile sources, with a random or fixed output.

Finally, for the last item we already started to design a competitive setting where multiple VOs, with different functionalities and reconfiguration strategies, compete for a set of limited resources. The quality of alternative reconfigurations strategies is tested by direct competition of VOs in the same environment. A more biologically inspired scene may be used, as well. In that scenario, VOs may compete ones against the others; in particular, a garbage collector need to be design, allowing to decompose dead VOs and reuse their resources (nodes). 

\paragraph{Related work}

Self-organizing and adapting systems appear in different areas of computer science. Robotics is one of the fields that intensively studied this paradigm \cite{yim2007modular, dorigo2004evolving, cao1997cooperative, hosokawa1999self, gilpin2010robot, wei2010sambot}. Modular self-reconfigurable robotic systems are studied in \cite{yim2007modular}. In applications such as space explorations, it is desirable to have a system that allows robots to configure their own shape, to adapt to new circumstances, or recover from damage. The authors are studying different architectures suitable for these systems, and one of the identified architecture is the ``Lattice Architecture'' which arranges modules in a regular grid similar to the one used in our paper. In \cite{dorigo2004evolving} the authors investigate methods for self-assembling and self-organizing systems to create the concept of ``swarm-bot'', which is defined as a composition of smaller autonomous robots; the authors are mainly concerned with the aggregation and coordinated motion of the robots for reaching certain goals such as path-finding or obstacle avoidance in open-space environments.

Self-organizing networks is another area of interest \cite{collier2004self, wischhof2005information, tang2003peer, cerpa2004ascent}. Self-configuring wireless networks are studied in \cite{collier2004self}. Considering a dynamic environment, limited power and low radio range, it becomes a real challenge to configure a well connected network using a small number of wireless devices. A solution for this challenge is a self-organization of devices using dynamic adaptation in a distributed environment.
Inter-vehicle communication for improving the passenger comfort and traffic efficiency is presented in \cite{wischhof2005information}. 
Data dissemination, information aggregation, and communication performance is studied in the context of ad-hoc wireless networks; for instance, in \cite{tang2003peer} a large scale peer-to-peer information retrieval system is designed, using a decentralized architecture.

Software engineering is another area where self-adaptive systems play a central role. In \cite{mckinley2004composing}, the authors identifies middleware as the layer developers can exploit to (compositionally) implement adaptive behavior. They also have identified three basic techniques for dealing with compositional adaptation of software systems: separation of concerns, computational reflexion, and component-based design. Separation of concerns is particularly used to get two distinct layer: one for system functionality, the other for its adaption. The latter usually involves a feedback loop, implementing MAPE (monitoring, analysis, planning, execution) cycle. In \cite{weyns2013patterns}, the authors describe patterns for decentralized control, used to cope with the complexity of this feedback loop. A protocol for decentralized self-assembly is presented in \cite{sykes2011flashmob}. A survey, including a broader view and also recent approaches, is presented in \cite{macias2013self}.

Formal methods are also applied to self-adaptive systems \cite{lewis2011survey, bruni2015modelling, khakpour2012formal, calinescu2015self, tamura2013towards}. A quantitative survey \cite{lewis2011survey} finds a relative low percent of formal methods papers from those dedicated to self-adaptive systems. One reason may be the difficulty to define what an adaptive system actually is \cite{bruni2015modelling}. Most of the papers are devoted to efficiency and reliability features, and less to functionality change, security, or scalability issues. From the modelling point of view, regular-expressions and automata are most frequently used models. Structural and functional adaptation are studied in \cite{khakpour2012formal}, exploiting a model based on actors, process algebra and policies for managing and adapting the system behavior. In \cite{bruni2015modelling} the authors use Maude to study adaptive systems, including  a case study dealing with robots' morphogenesis, obstacle avoidance, and collective healing. Verification and validation of self-adaptive systems are studied in \cite{tamura2013towards}, the authors emphasizing the need to evolve system behavior hand in hand with the evolution of the contracts they obey to. A recent paper \cite{calinescu2015self} is devoted to distributed control of self-adaptive systems. 

\paragraph*{Conclusion}
We expect that self-organizing and adaptive systems usage will grow and spread to many other domains and applications, considering the recent paradigms like ubiquitous or edge computing. However, decentralized system architectures and processes communicating within close range or receiving data from the external environment is a challenging field. We think that more architecture formalisms and programming languages are needed to prepare these type of systems for the future, our proposal being one going in that direction. 

\out{
\paragraph{Acknowledgements.} 
The research reported in this paper was partially supported by... 
}

\bibliographystyle{fundam}
\bibliography{bibtex_pad-stef}

\begin{thebibliography}{10}

\bibitem{abadi2016tensorflow}
Abadi, M., Agarwal, A., Barham, P., Brevdo, E., Chen, Z., Citro, C., Corrado,
  G.~S., Davis, A., Dean, J., Devin, M., et~al.: Tensorflow: Large-scale
  machine learning on heterogeneous distributed systems,
\newblock \emph{arXiv preprint arXiv:1603.04467}, 2016.

\bibitem{bps13}
Banu{-}Demergian, I., Paduraru, C., Stefanescu, G.: A new representation of
  two-dimensional patterns and applications to interactive programming,
\newblock \emph{FSEN 2013}, LNCS 8161, Springer, 2013.

\bibitem{ba-st14}
Banu{-}Demergian, I., Stefanescu, G.: Towards a Formal Representation of
  Interactive Systems,
\newblock \emph{Fundamenata Informaticae}, \textbf{131}, 2014, 313--336.

\bibitem{banu2016contour}
Banu-Demergian, I.~T., Stefanescu, G.: On contour representation of two
  dimensional patterns,
\newblock \emph{Carpathian Journal of Mathematics}, 2016, 37--47.

\bibitem{bruni2015modelling}
Bruni, R., Corradini, A., Gadducci, F., Lafuente, A.~L., Vandin, A.: Modelling
  and analyzing adaptive self-assembly strategies with Maude,
\newblock \emph{Science of Computer Programming}, \textbf{99}, 2015, 75--94.

\bibitem{calinescu2015self}
Calinescu, R., Gerasimou, S., Banks, A.: Self-adaptive Software with
  Decentralised Control Loops,
\newblock \emph{FASE}, 15, 2015.

\bibitem{cao1997cooperative}
Cao, Y.~U., Fukunaga, A.~S., Kahng, A.: Cooperative Mobile Robotics:
  Antecedents and Directions,
\newblock \emph{Autonomous Robots}, \textbf{4}, 1997, 7--27.

\bibitem{cerpa2004ascent}
Cerpa, A., Estrin, D.: ASCENT: Adaptive self-configuring sensor networks
  topologies,
\newblock \emph{IEEE transactions on mobile computing}, \textbf{3}(3), 2004,
  272--285.

\bibitem{collier2004self}
Collier, T.~C., Taylor, C.: Self-organization in sensor networks,
\newblock \emph{Journal of Parallel and Distributed Computing}, \textbf{64},
  2004, 866--873.

\bibitem{conway2012regular}
Conway, J.~H.: \emph{Regular algebra and finite machines},
\newblock Courier Corporation, 2012.

\bibitem{dennis1974first}
Dennis, J.~B.: First version of a data flow procedure language,
\newblock \emph{Programming Symposium}, Springer, 1974.

\bibitem{dijkstra1986derivation}
Dijkstra, E.~W., Feijen, W.~H., Van~Gasteren, A.~M.: Derivation of a
  termination detection algorithm for distributed computations,
\newblock in: \emph{Control Flow and Data Flow: concepts of distributed
  programming}, Springer, 1986,  507--512.

\bibitem{dorigo2004evolving}
Dorigo, M., Trianni, V., {\c{S}}ahin, E., Gro{\ss}, R., Labella, T.~H.,
  Baldassarre, G., Nolfi, S., Deneubourg, J.-L., Mondada, F., Floreano, D.,
  Gambardella, L.~M.: Evolving Self-Organizing Behaviors for a Swarm-Bot,
\newblock \emph{Autonomous Robots}, \textbf{17}, 2004, 223--245.

\bibitem{dr-st08a}
Dragoi, C., Stefanescu, G.: AGAPIA v0. 1: A programming language for
  interactive systems and its typing system,
\newblock \emph{Electronic Notes in Theoretical Computer Science},
  \textbf{203}(3), 2008, 69--94.

\bibitem{giammarresi1997two}
Giammarresi, D., Restivo, A.: Two-dimensional languages,
\newblock \emph{Handbook of formal languages}, \textbf{3}, 1997, 215--267.

\bibitem{gilpin2010robot}
Gilpin, K., Knaian, A., Rus, D.: Robot pebbles: One centimeter modules for
  programmable matter through self-disassembly,
\newblock \emph{Robotics and Automation (ICRA), 2010 IEEE International
  Conference on}, IEEE, 2010.

\bibitem{halbwachs1991synchronous}
Halbwachs, N., Caspi, P., Raymond, P., Pilaud, D.: The synchronous data flow
  programming language LUSTRE,
\newblock \emph{Proceedings of the IEEE}, \textbf{79}(9), 1991, 1305--1320.

\bibitem{hosokawa1999self}
Hosokawa, K., Fujii, T., Kaetsu, H., Asama, H., Kuroda, Y., Endo, I.:
  Self-organizing collective robots with morphogenesis in a vertical plane,
\newblock \emph{JSME International Journal Series C Mechanical Systems, Machine
  Elements and Manufacturing}, \textbf{42}, 1999, 195--202.

\bibitem{autonomic2003}
Kephart, J., D.M.Chess: The vision of autonomic computing,
\newblock \emph{Computer}, 2003.

\bibitem{khakpour2012formal}
Khakpour, N., Jalili, S., Talcott, C., Sirjani, M., Mousavi, M.: Formal
  modeling of evolving self-adaptive systems,
\newblock \emph{Science of Computer Programming}, \textbf{78}(1), 2012, 3--26.

\bibitem{kleene1951representation}
Kleene, S.~C.: \emph{Representation of events in nerve nets and finite
  automata},
\newblock Technical report, RAND PROJECT AIR FORCE SANTA MONICA CA, 1951.

\bibitem{lewis2011survey}
Lewis, P.~R., Chandra, A., Parsons, S., Robinson, E., Glette, K., Bahsoon, R.,
  Torresen, J., Yao, X.: A survey of self-awareness and its application in
  computing systems,
\newblock \emph{Self-Adaptive and Self-Organizing Systems Workshops (SASOW),
  2011 Fifth IEEE Conference on}, IEEE, 2011.

\bibitem{lindgren1998complexity}
Lindgren, K., Moore, C., Nordahl, M.: Complexity of two-dimensional patterns,
\newblock \emph{Journal of statistical physics}, \textbf{91}(5), 1998,
  909--951.

\bibitem{macias2013self}
Mac{\'\i}as-Escriv{\'a}, F.~D., Haber, R., Del~Toro, R., Hernandez, V.:
  Self-adaptive systems: A survey of current approaches, research challenges
  and applications,
\newblock \emph{Expert Systems with Applications}, \textbf{40}(18), 2013,
  7267--7279.

\bibitem{mckinley2004composing}
McKinley, P.~K., Sadjadi, S.~M., Kasten, E.~P., Cheng, B.~H.: Composing
  adaptive software,
\newblock \emph{Computer}, \textbf{37}(7), 2004, 56--64.

\bibitem{DBLP:conf/ispdc/Paduraru14}
Paduraru, C.: Dataflow Programming Using {AGAPIA},
\newblock \emph{Proceedings {ISPDC} 2014}, IEEE, 2014.

\bibitem{p15}
Paduraru, C.: \emph{Research on {AGAPIA} language, compiler and applications},
\newblock Ph.D. Thesis, University of Bucharest, 2015.

\bibitem{DBLP:conf/saso/PaduraruMS17}
Paduraru, C., Mincu, R., Stefanescu, G.: Multi-Level Control Mechanisms for
  Non-Structured and Structured 2-Dimensional Self-Assembling,
\newblock \emph{Proc. {SASO} 2017}, 2017.

\bibitem{ecsa}
Paduraru, C., Stefanescu, G.: Self-assembling heterogeneous interactive
  systems,
\newblock \emph{ECSA - Systems of systems 2016}, ACM, 2016.

\bibitem{pss07}
Popa, A., Sofronia, A., Stefanescu, G.: High-level Structured Interactive
  Programs with Registers and Voices,
\newblock \emph{J. {UCS}}, \textbf{13}, 2007, 1722--1754.

\bibitem{DBLP:conf/RelMiCS/Stefanescu15}
Stefanescu, G.: A Quest for {K}leene Algebra in 2 Dimensions,
\newblock \emph{Proceedings {RAMiCS} 2015}, LNCS, Springer, 2015.

\bibitem{stoica2001chord}
Stoica, I., Morris, R.~T., Karger, D.~R., Kaashoek, M.~F., Balakrishnan, H.:
  Chord: {A} scalable peer-to-peer lookup service for internet applications,
\newblock \emph{{SIGCOMM}}, 2001.

\bibitem{sykes2011flashmob}
Sykes, D., Magee, J., Kramer, J.: Flashmob: distributed adaptive self-assembly,
\newblock \emph{Proceedings of the 6th International Symposium on Software
  Engineering for Adaptive and Self-Managing Systems}, ACM, 2011.

\bibitem{tamura2013towards}
Tamura, G., Villegas, N., M{\"u}ller, H., Sousa, J.~P., Becker, B., Pezze, M.,
  Karsai, G., Mankovskii, S., Sch{\"a}fer, W., Tahvildari, L., et~al.: Towards
  practical runtime verification and validation of self-adaptive software
  systems, 2013.

\bibitem{tang2003peer}
Tang, C., Xu, Z., Dwarkadas, S.: Peer-to-peer Information Retrieval Using
  Self-organizing Semantic Overlay Networks,
\newblock \emph{Proceedings {SIGCOMM} '03}, ACM, 2003.

\bibitem{wei2010sambot}
Wei, H., Cai, Y., Li, H., Li, D., Wang, T.: Sambot: A self-assembly modular
  robot for swarm robot,
\newblock \emph{Robotics and Automation (ICRA), 2010 IEEE International
  Conference on}, IEEE, 2010.

\bibitem{weyns2013patterns}
Weyns, D., Schmerl, B., Grassi, V., Malek, S., Mirandola, R., Prehofer, C.,
  Wuttke, J., Andersson, J., Giese, H., G{\"o}schka, K.~M.: On patterns for
  decentralized control in self-adaptive systems,
\newblock in: \emph{Software Engineering for Self-Adaptive Systems II},
  Springer, 2013,  76--107.

\bibitem{wischhof2005information}
Wischhof, L., Ebner, A., Rohling, H.: Information dissemination in
  self-organizing intervehicle networks,
\newblock \emph{IEEE Transactions on intelligent transportation systems},
  \textbf{6}, 2005, 90--101.

\bibitem{wolfram2002new}
Wolfram, S.: \emph{A new kind of science}, vol.~5,
\newblock Wolfram media Champaign, 2002.

\bibitem{yim2007modular}
Yim, M., Shen, W.-M., Salemi, B., Rus, D., Moll, M., Lipson, H., Klavins, E.,
  Chirikjian, G.~S.: Modular self-reconfigurable robot systems [grand
  challenges of robotics],
\newblock \emph{IEEE Robotics \& Automation Magazine}, \textbf{14}(1), 2007,
  43--52.

\end{thebibliography}

\section{Appendix: 2D regular expressions}\label{ap-a}

This appendix presents a few details on using regular expressions to define 2D languages. The regular expressions included are related to the $RAT$ language (Rings with attached trees), defined in Section \ref{sec2}. 

\subsection{Regular 2D expressions} 

A (horizontally-vertically) connected 2D word is represented by specifying its shape and the labels of the included cells. As such a connected 2D word may have internal holes, its shape is specified with a \textit{normal-form contour (nf-contour)} \cite{banu2016contour}, collecting the external contour of the word and the word's holes contours, namely $$(C_0; (C_1,x_1,y_1),\dots,(C_k,x_k,y_k))$$  where:
\begin{itemize}
\item $C_0$ is the external contour of the word, starting with $P_0$ and recording the border elements going clockwise, where $P_0$ is the left-most corner of the top row of the word;
\item $k$ is the number of the holes; 
\item each $(C_i,x_i,y_i)$, $i\in\{1,\dots,k\}$, describes an internal hole, where $(x_i,y_i)$ is the offset of its starting point $P_i$ (the left-most corner of the top row of the hole) and $C_i$ is the contour of the hole, starting with $P_i$ are recording the hole's borders in the counterclockwise order;
\item to have a unique representation, the holes are ordering using the lexicographical order of their offsets $(x_i,y_i)$. 
\end{itemize}
With this representation, each point on the word's border is represented as $i.n$, where $i\in\{0,\dots,k\}$ specifies the contour component of the point and $n$ is the position of the point in that contour, counted starting with $0$. Similarly, an interval on the word's border is represented as $i.[m,n]$ and consists of the edges of the contour $i$, placed in between points $i.m$ and $i.n$. We also use the extended notations $v:i.n$ and $v:i.[m,n]$, including a reference to the word to which the point or interval border belongs.

\begin{figure}[tbh]\begin{center}
\begin{tabular}{c@{\hspace{2cm}}c}
\includegraphics[scale=0.3]{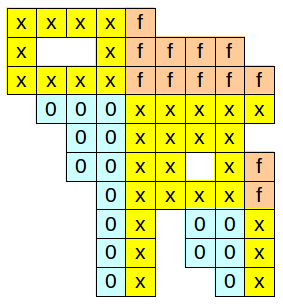}
&\includegraphics[scale=0.2]{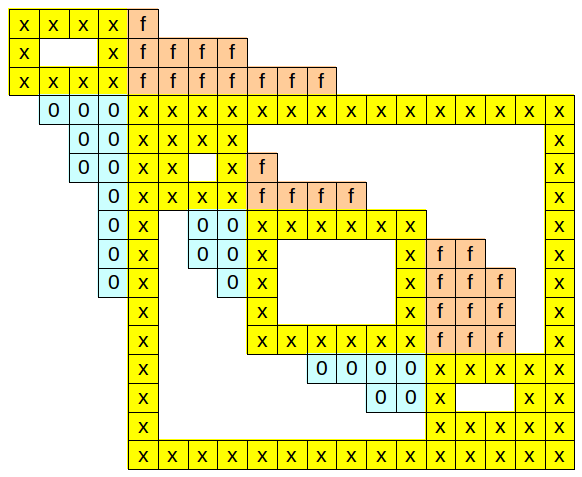}
\\(a)&(b)\vspace{-.5cm}
\end{tabular}\end{center}
\caption{2D words (with holes)}\label{f-words}
\end{figure}

For example, the contour of the word in Fig.~\ref{f-words}(a) is represented as
\svsp\\ 
\centerline{$(r^5dr^3drd^2ldrd^5l^2ulu^2ld^3l^2u^4lu^2lulu^3;  (dr^2ul^2,1,1), (drul,6,5))$.}\svsp\\
The top-right corner in the small hole is represented as the point $2.3$, while the bottom-left corner of the left-most $0$-cell is $0.41$ (it may be easier to count in the opposite order, using negative numbers; then the point is $0.\!\!-\!\!5$). The interval $0.[4,12]$ refers to the external border of the top group of $f$-cells.  

With this more detailed representation of the contours, the 2D word composition can be controlled with a finer granularity. For instance, the general 2D word composition operator \cite{DBLP:conf/RelMiCS/Stefanescu15} may be more formally expressed with this notation. Indeed: let $v1$ and $v2$ be two words; let $P1$ (resp. $E1$) be a set of points (resp. length 1 intervals) on the contour of $v1$ and similarly $P2$ (resp. $E2$) for $v2$; let $\rho$ be a comparison relation between them (for instance, use equality `$=$', inclusion `$<$', or nonempty intersection `$\#$') and denote by $P1\ \rho\ P2$ (resp. $E1\ \rho\ E2$) the resulting atomic formula; finally, let $\phi$ denote a boolean formula built up with these atomic formulas; then, the general 2D word composition operator is $v1\ \phi\ v2$, producing all non-overlapping arrangements of $v1$ and $v2$ such that the relation $\phi$ between the selected points and length 1 intervals on the contours of $v1$ and $v2$ is valid.

Two examples of 2D word composition instances, using this notation, are:
\begin{itemize}
\item $v1\ (3.5 = 0.2)\ v2$ specifies a composite of the words $v1$ and $v2$, where the $5$-th point on the contour of the $3$-th hole of $v1$ is identified with the $2$-nd point of the external contour of $v2$, both points being uniquely identified according to the nf-contour representations of $v1$ and $v2$. (Notice that the result of this composition is unique, provided the words do not overlap after these points' identification.) 
\item Similarly, $v1\ (2.[3,7] < 0.[10,20])\ v2$ say that after the composition, the interval $[3,7]$ on the $2$-nd hole of $v1$ is included in the interval $[10,20]$ on the external contour of $v2$. (In this case, in principle, this composition may produce more results, depending on the way the interval $v1:2.[3,7]$ is included in the interval $v2:0.[10,20]$.)
\end{itemize}

This simplified formalism for defining general 2D word composition operators easily captures the original setting for defining the class n2RE of new 2D regular expressions \cite{bps13}, \cite{ba-st14}. Indeed, an nf-contour capture the needed information to say when a point $i.n$ is $nw$-, $ne$-, $se$-, or $sw$- land or golf corner, and when a length 1 interval $i.[n,n+1]$ is a west, north, east, or south edge border. For instance, a point $i.n$ is $se$-golf corner if it occurs in between a `$d$' and an `$l$' in the contour, $d$ being a west and $l$ a  north side of the word. The current alternative of specifying border elements is more precise, but requires a global view on a word, while the former version for n2RE uses a local view and fit better when composition is applied to sets of words.

Finally, we introduce two useful notations:\bi
\item near-k-nw - a border point obeys this restriction if it is at distance k of a nw-corner; for a border edge, the condition means one of its end points  satisfies it.  
\item strict(C) - strict, in front of a condition C, means: ``except for the contact of border elements specified in C, no other border elements on the composing words are shared in the composed word''.
\ei

\subsection{A regular expression for the RAT language}

The difficult part in describing the RAT language is to generate membranes compositionally. These membranes are minimal words, in the sense one cannot eliminate cells, remaining in the language. Every word in the RAT language is obtained starting with such a membrane and recursively connecting bars of 2's, 4's, 7's, and e's, with their north, east, south, and west sides, respectively.

\begin{figure}[h]
\centerline{
\raisebox{.2cm}{\includegraphics[scale=0.25]{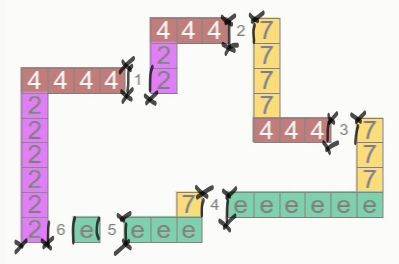}}
\includegraphics[scale=0.25]{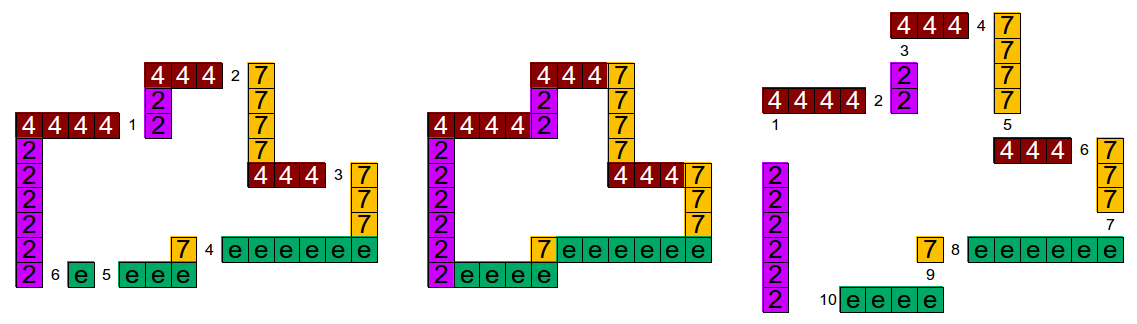}
}
\caption{A ring (minimal 2D-word in the RAT-language), in the 3rd position, and two decompositions (left-right) for describing possible generations with regular 2D-expressions. In front, an illustration of the chosen border elements for word compositions, following the decomposition in the paper and illustrated in the 2nd picture.}\label{nf9}\vspace{-.5cm}
\end{figure}

To generate membranes, in the clockwise or anticlockwise order, we will use a few intermediary expressions with the following intuitive meanings: 
\bi\item 
E2 generates columns of 2's; E7 columns of 7's; E4 rows of 4's; Ee rows of e's; 
\item F24 generates corners consisting of a column of 2's and a row of 4's; similarly, F2e, F74, F7e;
\item G1 generates chains of connected corners and having only one horizontal end: either (1) to the right (the end of a row of 4's), identified by the condition ``e \& near-0-ne \& near-0-se''; or (2) to the left (the end of a row of e's), identified by the condition ``w \& near-0-nw \& near-0-sw''. To these horizontal ends, we freely connect appropriate new 24-, 2e-, 74-, or 7e-corners;
\item The first alternative in G2 asks to connect the last e-end to the other end of the chain (the bottom of the starting column of 2's). To identify this position, we use the condition ``e \& near-0-se \& near-1-sw''. (For instance, in the illustrated figure, without the latter condition, there are two candidates satisfying the ``e \& near-0-se'' restriction.). The second alternative is similar, describing the closing of the cycle with a 4's bar.
\item Rat expression allows to recursively connect bars of 2's, 4's, 7's, and e's to G2.
\item finally, the compositions in G1, G2, Rat are \textit{strict}, in the sense, except for contact border elements specified by the restriction, no other connections of border elements are allowed.
\ei
These expressions are formally defined as follows: 
\be
\item E2 = 2 + 2 (s=n) E2;
\item E4 = 4 + 4 (e=w) E4;
\item E7 = 7 + 7 (s=n) E7;
\item Ee = e + e (e=w) Ee;
\item F24 = E2 (n = (s \& near-0-sw)) E4
\item F2e = E2 (n = (s \& near-0-se)) E4
\item F74 = E7 (s = (n \& near-0-nw)) E4
\item F7e = E7 (s = (n \& near-0-ne)) Ee
\item G1 = F24 + F2e
\\+ G1 strict((e \& near-0-ne \& near-0-se) = (w \& near-0-sw)) (F24 + F2e)
\\+ G1 strict((e \& near-0-ne \& near-0-se) = (w \& near-0-nw)) (F74 + F7e)
\\+ G1 strict((w \& near-0-nw \& near-0-sw) = (e \& near-0-ne)) (F24 + F2e)
\\+ G1 strict((w \& near-0-nw \& near-0-sw) = (e \& near-0-ne)) (F74 + F7e)
\item G2 = G1 strict(((w \& near-0-nw \& near-0-sw) =  e)
\& ((e \& near-0-se \& near-1-sw) = w)) Ee
\\+ G1 strict(((e \& near-0-ne \& near-0-se) =  w) 
\& ((w \& near-0-sw \& near-1-se) = e)) E4
\item Rat = G2 
\\+ E2 strict(n \texttt{<} s) RAT + E4 strict(e \texttt{<} w) RAT 
\\+ E7 strict(s \texttt{<} n) RAT + Ee strict(w \texttt{<} e) RAT
\ee

\paragraph{Remark 1.}
A first, simpler attempt is to connect bars of 2's, 4's, 7's, and e's, in the order 1,...,10 described in the right part of Fig.~\ref{nf9}. The approach fails,  as in connection 5 we cannot discriminate between the two ends of the chain (the south of the starting bar of 2's and the one of 7's in link 5), both in the south direction. The 2D automata, describing this language, shows a composition is needed to cover the south of a 7-cell, which is not an accepted border, but it is not needed for the south border of a 2-cell.

\paragraph{Remark 2.}
The obtained expression Rat is a bit complicate as we intended to show how to describe a cycle (membrane), without any tree attached to it. For the full $RAT$ language, probably a simpler expression may be found.

\section{Appendix: Implementing virtual organisms in AGAPIA}\label{ap-b}

Agapia is a DSL structured programming language for interactive systems  \cite{dr-st08a,pss07,DBLP:conf/ispdc/Paduraru14,p15}.  
The current version of its compiler allows quick design of HPC applications deployed using MPI and OpenMP platforms. It is open source and available at \svsp\\\centerline{\url{https://github.com/AGAPIA/CompilerAndTools}}\svsp\\ 
The package contains a manual on how to use it, and sample problems implemented and ready to run.
The starting point for building Agapia programs are interactive modules. Modules have data on interfaces, but no labels to allow for control and interaction coordination patterns. Traditionally, a program is seen as a transformation with input data on the west-north interfaces and output data at the east-south interfaces. 

While AGAPIA promotes a loosely-coupled architecture, it is also able to impose restrictions over the communicating components in distributed systems. The default communication model is directly between sender and receiver, instead of using a centralized node (such as a broker). In consequence, this architectural decision resolves the message overloading problem in the publish-subscriber paradigm. However, inside programs users are free to mix between the two paradigms.
In applications composed of several code pieces it is important nowadays to provide ways to parallelize the execution process. AGAPIA provides a background parallel execution of programs in a transparent way, i.e., user does not write explicit parallel work for creation or synchronization. The runtime component of AGAPIA tries to maximize the parallelism inside an application by mapping ready for execution instances of programs (instances with no remaining input dependencies) to workers: threads - when being executed in the shared memory model, or processes on different machines - when executed in the distributed memory model.

Regarding performance, as evaluated in \cite{p15}, the only overhead of the AGAPIA language compared to a hand-written application in a low-level programming language comes from communicating data between modules when performing interpretation. This internal overhead was evaluated at less than $7\%$ for the shared memory model and around $15\%$ for the distributed execution model. It is important to state that the C/C++ code inside modules is still compiled and added as native code in the final binary file. The only interpreted part of an application written in AGAPIA are its high-level statements and composition operators that usually drive the execution of the native code. Overall, the implementations in AGAPIA can prove to be faster to write, maintain and extend, less error prone at a small performance incur.

We have tried to estimate the users' requirements from a language that supports virtual organisms, and came up with the following list:
\begin{itemize}
\item Specification of a virtual organism structure, as input.
\item Allow user to inject custom functionality (callbacks) for its cells.
\item Deploy it on a distributed system to support different applications (e.g., such as the ones in the IoT domain).
\end{itemize}

In this section, we briefly describe how Agapia allows the implementation and deployment of virtual organisms - see the indicated references for more details related to the current syntax and semantics of the language.

\subsection{Constructing a virtual organism}

The user can specify a virtual organism by one of the following three methods:
\begin{enumerate}
\item Plain text specifying the cell names on lines and columns, and using the symbol ``*'' to denote an empty space.
\item Specify a 2D regular expression to create a grid of cells using the product of the languages.
\item Specify a C/C++ functor to the AGAPIA's initialization runtime that creates the grid of cells using a custom representation. This lets users specify for each cell which are its children and parent.
\end{enumerate}

For example, the $TC$ model structure can be specified using method 2, while for describing models $RAT$, $CRAT$ and other custom based virtual organisms, method 3 can be used.

\begin{figure*}
\centerline{\includegraphics[scale=0.7]{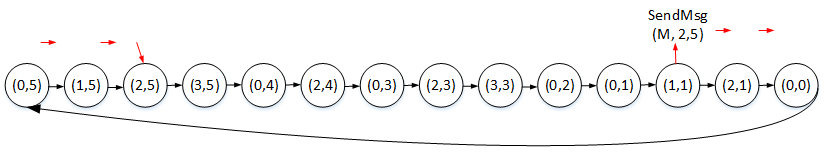}}
\vspace{-.3cm}
\caption{The ring of processes created after linearizing the structure from Figure 3. With red labels, it is shown the path of a message from the cell located at (1,1) to (0,5).}\label{AGAPIA_TCex2}
\end{figure*}

\subsection{How can user inject custom functionality}

After creating the structure, it is desirable to be able to customize the behavior of the cells in the virtual organism. To do so, we reuse the AGAPIA's concept of module definition. Typically, a module definition per cell symbol type in the source code must exist, otherwise its behavior is considered an identity. At runtime, a module instance is built for all existing cells (i.e., if a symbol appears $N$ times in the grid, there will be $N$ modules instances of the corresponding type). The pseudocode in Listing \ref{TCpseudocode} sketches the $TC$ model defined in Section \ref{subs:reconfig}. As noticed in the code, to facilitate the development of virtual organisms, there are a few language constructs added in the syntax of each module's code (and automatically bound per instance).

Thus, each cell instance:
\begin{itemize}
\item Has members to get the parent and children list:
\\ ($member.Parent,  member.ChildrenList$)
\item Has a coordinate in the grid - created internally at construction step: 
\\ $member.row, member.col$
\item Knows if it is a leaf or root: $member.isLeaf, member.isRoot$. This avoids code duplication in the case of modules defining the same cell type but some of them being a leaf, for example.
\item Can send data to a different cell in the grid by knowing its coordinate.
\item Has a message box with messages received from other cells in the grid.
\end{itemize}

\begin{lstlisting}[mathescape,caption=Pseudocode skeleton for the TC model described in section \ref{subs:reconfig}, label=TCpseudocode, 
basicstyle=\small\ttfamily,basewidth  = {.5em,0.3em}]
module 4{listen nil}{read nil}
{
 while(true)
 {
  // Capture flow if this is a leaf node
  DataFlow df = {0};
  if (this.isLeaf)
     df = captureFromLeafsAround()
  
  // Process the messages existing in 
  // the message box
  Message M;
  while(M = this.MessageBox.Pop())
  {
    // If flow message received from children
    // add it to the local variable
    if (M.type == FLOW)
       df += (DataFlow) M;

    // if reorganization message received, follow
    // the algorithm defined in section 4.1
    if (M.type == REORG)
      DoReorganization();
   }
 
  // Process captured data flow (df) according
  // to the custom users' rules
  ProcessCapturedFlow(df);

  // Send flow further to this instance parent
  SendMsg(df, parent.Row, parent.Col);
 }
}{speak nil}{write nil}

module 2{listen nil}{read nil}
{
 // Possibly similar code to module 4
 // Excepting that we have a different process 
 // method for transforming the captured data 
}{speak nil}{write nil}

module 6{listen nil}{read nil}
{
 while(true)
 {
  // From time to time or by a given user
  // algorithm call reorganization
  // in the entire organism
   if (...)
     Broadcast(REORG)
  
 // Get the dataflow from MessageBox 
 // and process it  similar to the definition 
 // of module 4
 }
}{speak nil}{write nil}
\end{lstlisting}

\subsection{Implementation details}

Since the virtual organisms structures can be very different depending on user input and method used, the approach that we came up to make the implementation easier (more generic),  less error prone and without sacrificing the asymptotic communication complexity, was to create a ring topology that contains all the organism's cells. The user can control the order of the nodes in the ring with one of the two methods:

\begin{itemize} 
\item Using one of the default methods implemented: right to left or left to right for columns, and bottom to top or top to bottom for rows.
\item Using a custom user functor given as a parameter to the system initialization that given the grid of cells defines how to collect them in the output ring.
\end{itemize}

\begin{figure}
\centerline{\includegraphics[scale=0.8]{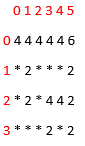}}
\vspace{-.3cm}
\caption{An example of user defined TC-organism}\label{AGAPIA_TCex}
\end{figure}

Continuing the exemplification on the $TC$ example, and supposing that a user defined an organism structure similar to the one in Figure \ref{AGAPIA_TCex}, then its ring structure - in order of right to left columns and top to bottom rows -  can be visualized in Figure \ref{AGAPIA_TCex2}. The method used for selecting the order can influence the communication time. While asymptotic complexity remains $O(N)$, where $N$ is the number of cells in the ring, choosing a bottom rows order method would decrease the communication effort needed in the $TC$ problem. This is one of the reasons why we let users provide their own functor and have fine control over the process.

At the implementation level, as seen in Listing \ref{TCpseudocode}, each cell has a MessageBox member. Messages flow from cell to cell in the ring until they reach their destination. Considering a distributed system with $P$ processes, at the deployment level it is possible that more than one cell to be assigned to the same process. Our current strategy is to split the ring of nodes in $P$ equal chunks of cells (group), and assign contiguous chunks of nodes to the same process. Thus, the real communication happens only between groups of cells and the message boxes per each cell are entries in a data structure that holds all message boxes in its parent group. Each process has two threads: one that handles communication (and sleeps when no communication is needed), and another one that processes the owned group of cells.

\end{document}